\documentclass[fleqn,usenatbib]{mnras}

\usepackage{newtxtext,newtxmath}

\usepackage[T1]{fontenc}

\DeclareRobustCommand{\VAN}[3]{#2}
\let\VANthebibliography\thebibliography
\def\thebibliography{\DeclareRobustCommand{\VAN}[3]{##3}\VANthebibliography}

\usepackage{graphicx}	
\usepackage{amsmath}	
\usepackage[para,online,flushleft]{threeparttable}

\newcommand{\obj}{NGC~4293}

\title[Low-mass AGN in \obj]{A compact symmetric ejection from the low mass AGN in the LINER galaxy NGC~4293}

\author[X. Yang et al.]{
Xiaolong Yang,$^{1,2}$\thanks{E-mail: yangxl@shao.ac.cn}
Ruiling Wang,$^{3}$
Quan Guo$^{1}$
\\
$^{1}$Shanghai Astronomical Observatory, Key Laboratory of Radio Astronomy, Chinese Academy of Sciences, Shanghai 200030, China\\
$^{2}$Shanghai Key Laboratory of Space Navigation and Positioning Techniques, Shanghai Astronomical Observatory, Chinese Academy of Sciences, Shanghai 200030, China\\
$^{3}$Astronomical Research Station of Shanghai Astronomical Observatory, Chinese Academy of Sciences, Shanghai 200030, China
}

\date{Accepted XXX. Received YYY; in original form ZZZ}

\pubyear{2022}

\begin{document}
\label{firstpage}
\pagerange{\pageref{firstpage}--\pageref{lastpage}}
\maketitle

\begin{abstract}
We conducted a Very Long Baseline Array (VLBA) observation of the low mass active galactic nucleus (AGN) in galaxy NGC~4293 ($z=0.003$). The object is associated with a low-ionization nuclear emission-line region (LINER). Its black hole mass is estimated as $\sim10^5$ or $\sim10^7 M_\odot$. The VLBA 1.5\,GHz image shows an inverse symmetric structure with two discrete radio blobs separated by an angular distance of $\sim120$\,mas, corresponding to $\sim7$\,parsec. Furthermore, its integrated radio spectrum has a turnover at the frequency of $\sim0.44$\,GHz. Based on the compactness and spectrum, the nuclear radio source in NGC~4293 belongs to a sample of (megahertz) peaked spectrum (PS/MPS) radio sources with compact symmetric morphologies. \obj\ has 1.4\,GHz radio power of only $\sim10^{20}\,\mathrm{W\,Hz^{-1}}$ with the VLBA observation, which is consistent with local AGNs but lower than the current PS samples. One of the two blobs has a steep radio spectrum $\alpha=-0.62\pm0.08$ ($S_\nu\propto\nu^{+\alpha}$), while the other one has an inverted spectrum $\alpha=0.32\pm0.10$. The VLBA 1.5\,GHz luminosity ratio of the two blobs is 3.23 and both blobs show lateral-flowing structures where the hotspots reside at the edge of each radio lobe. This can be explained as jet interactions with dense circumnuclear medium. We estimate the black hole mass of \obj\ through the fundamental plane of black hole activity, which constrains the black hole mass to be $\lesssim10^6\,M_\odot$. It supports that the object is a low-mass AGN and a potential candidate for accreting and ejecting IMBHs. 
\end{abstract}

\begin{keywords}
galaxies: active -- galaxies: individual: \obj\ -- radio continuum: galaxies -- galaxies: jets
\end{keywords}



\section{Introduction\label{sec:intro}}
Stellar-mass black holes (SBHs), formed from the direct collapses of massive stars \citep{2017NewAR..78....1M}, are widely observed in the Universe. Supermassive black holes (SMBHs, $10^6-10^{10}\,M_\odot$) are also universally found in the centers of galaxies with bulges \citep{2013ARA&A..51..511K}. However, the gap between two classes of black holes, i.e. the so-called intermediate-mass black holes \citep[IMBHs, $10^2-10^6\, M_\odot$, see review by][]{2020ARA&A..58..257G}, remains unfilled (especially $10^3-10^4\, M_\odot$). The necessity for the existence of IMBHs is strongly suggested by SMBHs with masses up to $10^{10}\,M_\odot$ \citep{2015Natur.518..512W, 2018Natur.553..473B} in the early Universe, when the Universe was only $5\%$ of its current age. These discoveries pose challenges to the formation of SMBHs, and intermediate-mass (seed) black holes are required to form SMBHs at such an early age.

Based on the expected formation and evolution scenarios, observational searches for IMBHs have typically focused on the following habitats: globular clusters, ultra/hyper-luminous X-ray sources, and dwarf galaxies, there is growing evidence for the abundance of $10^5\,M_\odot$ IMBHs in dwarf galaxies with stellar masses ranging from $10^7-10^{10}\,M_\odot$ \citep[e.g.][]{2004ApJ...610..722G, 2007ApJ...670...92G, 2012ApJ...755..167D, 2013ApJ...775..116R, 2018ApJS..235...40L, 2018ApJ...863....1C, 2018MNRAS.478.2576M}. Moreover, black holes with $10^4-10^6$ solar masses discovered in the centers of nearby dwarf galaxies are thought to be formed in the early Universe \citep{2010A&ARv..18..279V,2012NatCo...3.1304G}. These dwarf galaxies have undergone few merger events in their evolutionary history and therefore have not grown significantly since their birth. Thus, the central black holes in dwarf galaxies still retain the footprints of the seed black holes, providing valuable clues for our study of the formation and evolution of SMBHs in the early Universe \citep{2010A&ARv..18..279V,2020ARA&A..58...27I}.

Radio jets/outflows are the fundamental parts of both SBHs and SMBHs which maintain accretion process \cite[e.g.][]{2012MNRAS.425..605F,2015ARA&A..53..115K}. Continuum imaging and spectral studies can be used to infer the observational signatures of mildly relativistic to relativistic outflows \cite[0.1 -- 1.0 $c$ respectively, e.g.][]{2017MNRAS.464.1029N,2018A&A...616A.152S,2018MNRAS.477..830H,2021MNRAS.500.2620Y} and hence, the activity of IMBHs that are hosted in Active Galactic Nuclei (AGNs). A pilot radio survey and study of low-mass AGNs with high accretion rates by \citet[][]{2006ApJ...636...56G} indicates that these sources are predominantly radio-quiet. These signatures may be explained by comparison with Galactic X-ray binaries (hosting stellar mass black holes) where the high/soft X-ray state is characterized by a quenched radio emission \cite[e.g.][]{2006csxs.book..157M}. It is generally believed that the accretion process in SMBHs and SBHs are similar and scale independent \citep[e.g.][]{2017A&A...603A.127S}. Simultaneously, the high/soft state in Galactic X-ray binaries is radio quiet, while the transition from a low/hard state to a high/soft state is often associated with transient jets/outflows. On studying of a sample of high accretion rate IMBHs show that the ejection process in IMBHs is likely episodic, and in some cases the radio core is visible and in other cases, it is not (Yang et al. 2022, In preparation). A recent study of the IMBH candidate NGC~4395 reveals both fossil radio ejecta and diffuse and flat-spectrum radio emission in the nucleus \citep[][]{2022MNRAS.514.6215Y}, and a thermal mechanism in the nucleus is proposed correspondingly.

Very Long Baseline Interferometry (VLBI) observations offer high angular resolutions (at the milli-arcsec scale) in radio band, surpassing other imaging techniques in astronomy. The VLBI detection of compact pc-scale radio-emitting structures (core/core-jet/jet-knot) in the nuclear regions of dwarf galaxies can directly probe the jet/outflow activity enabled by an accreting, potential IMBH. To date, high-resolution VLBI observational studies are limited to only a few IMBH candidate hosts: NGC~4395 \citep[][]{2006ApJ...646L..95W,2022MNRAS.514.6215Y}, Henize~2-10 \citep[][]{2012ApJ...750L..24R}, NGC~404 \citep[][]{2014ApJ...791....2P} and RGG~9 \citep[][]{2020MNRAS.495L..71Y}. In order to understand the ejection process in actively accreting IMBHs, we present an observational study of an IMBH candidate in NGC~4293.

NGC~4293, at redshift $z=0.003$, was identified as a late-type spiral galaxy with a low-ionization nuclear emission-line region \citep[LINER,][]{1997ApJS..112..315H, 2007MNRAS.381..136D}. The black hole mass of its AGN was estimated as $\log{(M_\mathrm{BH}/M_\odot)}=7.7$ through the black hole mass ($M_\mathrm{BH}$) and velocity dispersion ($\sigma$) relation \citep{2005ApJ...625..716C}. Another work obtained a similar black hole mass $\log{(M_\mathrm{BH}/M_\odot)}=7.5$ by taking the black hole mass - $K_s$-band bulge luminosity relation \citep{2006AJ....131.1236D}. Recently, \citet{2018ApJ...863....1C} used the single-epoch broad H$\alpha$ emission to measure the black hole mass of \obj. This source was re-weighted as $\log{(M_\mathrm{BH}/M_\odot)}=5.30\pm0.06$, and hence an IMBH candidate. \citet{2002A&A...392...53N,2005A&A...435..521N} observed \obj\ at 15\,GHz with Very Large Array (VLA) A-array, they obtained a resolution of 0.15\,arcsec and yielded a integrated flux density of $\sim1.4$\,mJy (with merely a signal-to-noise ratio of $\sim5$). With the VLA A-array 15\,GHz observation and VLA archive data at 1.5 and 5\,GHz, \citet{2002A&A...392...53N} measured a steep radio spectrum ($\alpha<-0.3$ according to their definition and $S\propto\nu^{+\alpha}$, while in this work, a spectrum is termed as steep, flat or inverted when $\alpha<=-0.5$, $>-0.5$ or $>0$, respectively) in \obj. While comparing with the 15\,GHz observation and a Multi-Element Radio Linked Interferometer Network (MERLIN) 5\,GHz observation, \citet{2006A&A...451...71F} conclude a flat radio spectrum. In order to investigate the nuclear (parsec and sub-parsec scale) outflows of \obj, in this work, we report the Very Long Baseline Array (VLBA) observation of the nuclear region in NGC~4293 and we also analyse the archival data from VLA and MERLIN. This paper is organised as follows. In Section \ref{sec:observations}, we describe multi-band observations and data reduction of the target \obj. In Section \ref{sec:results}, we present imaging, spectral and model-fitting results of \obj. In Section \ref{sec:dis}, we discuss the radio identification of \obj, describe the interpretation of the multi-band properties, and estimate black hole mass. Finally, we give our conclusions in Section \ref{sec:conclusions}. Throughout this work, we adopt the standard $\Lambda$CDM cosmology with $H_0 =71$ km s$^{-1}$ Mpc$^{-1}$, $\Omega_\Lambda=0.73$, $\Omega_m=0.27$ and the corresponding physical scale is 0.061\,pc\,mas$^{-1}$ in \obj.

\section{Observation and data reduction} \label{sec:observations}
We observed \obj\ with eight VLBA antennas from 2021 July 08 to 09 (UT, the project ID: BA146). The antennas at Kitt Peak (KP) and North Liberty (NL) were unavailable due to equipment failure. The observation was scheduled at L-band (the central frequency is 1.545\,GHz, hereinafter using 1.5\,GHz for short), with a total observation time of 3 h and a data recording rate of 2\,Gbits per second. Phase-reference mode was used and the quasar 1216$+$179 (R.A. $=\mathrm{12^{h}18^{m}46^{s}.6045}$, DEC. $=+17^{\circ}38^{\prime}17^{\prime\prime}.268$) was chosen as the phase-reference calibrator. The observation was scheduled with a dual polarization mode and a total bandwidth of 320\,MHz with 1\,min on calibrator and 5\,min on target in each observing cycle. The correlated data was processed using the Astronomical Image Processing System \citep[AIPS,][]{2003ASSL..285..109G} developed by the National Radio Astronomy Observatory (NRAO) of the USA. A prior amplitude calibration was performed using the system temperatures and the antenna gain curves provided by each VLBA station. The Earth orientation parameters were obtained and calibrated using the measurements from the U.S. Naval Observatory database, and the ionospheric dispersive delays were corrected from a map of the total electron content provided by the Crustal Dynamics Data Information System (CDDIS) of NASA \footnote{\url{https://cddis.nasa.gov}}. The opacity and parallactic angles were also corrected using the auxiliary files attached to the data. A global fringe-fitting on the phase-reference calibrator 1216$+$179 was performed, taking the calibrator's model to solve miscellaneous phase delays of the target.

\begin{figure}
\centering \includegraphics[scale=0.6]{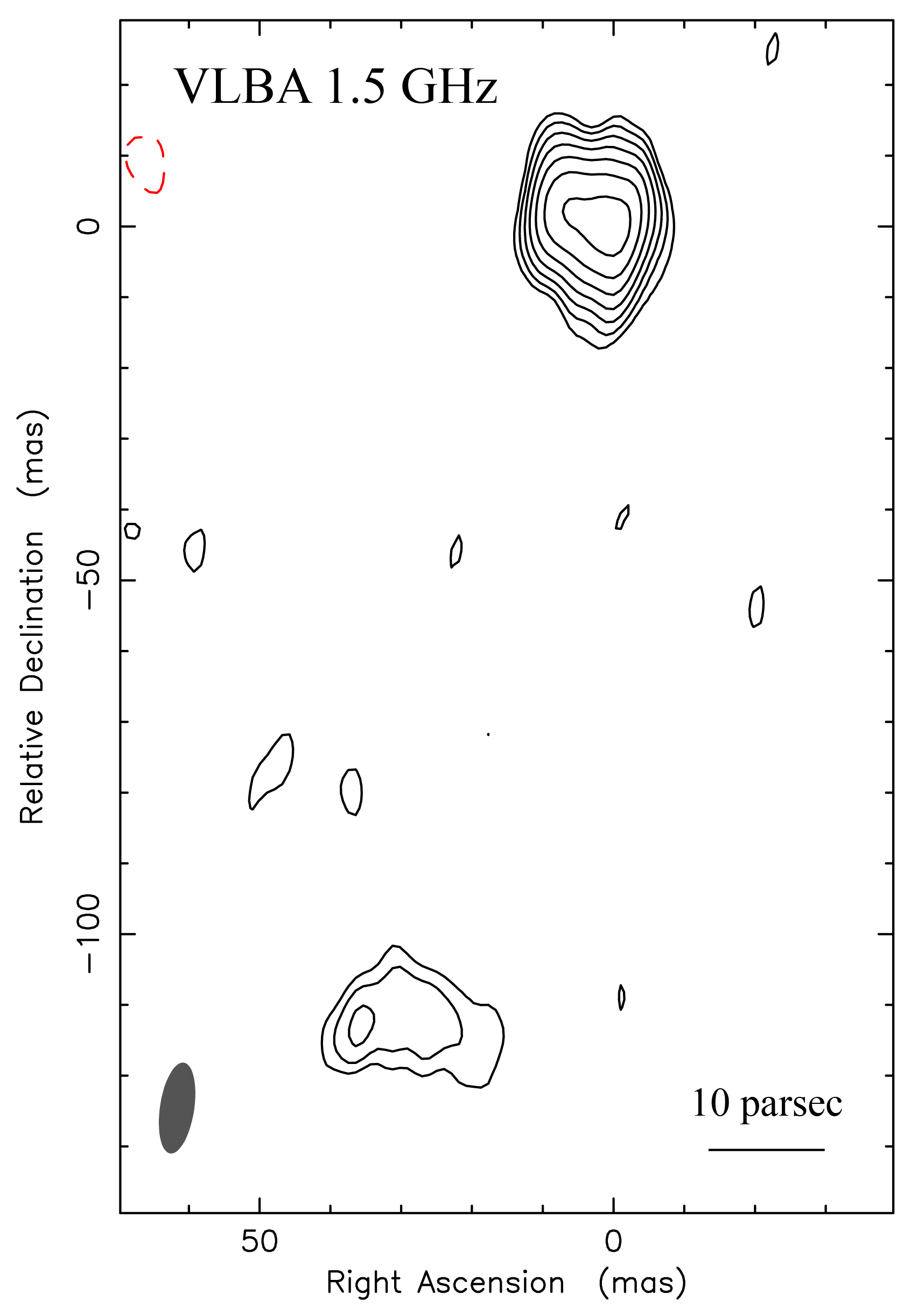}
\caption{Naturally-weighted VLBA 1.5\,GHz image of the core region in \obj. The map reference is at the peak position of radio flux density. The contours are at 3$\sigma\times(-1, 1, 1.41, 2, 2.83,...)$, here $1\sigma$ noise is $0.03$\,mJy\,beam$^{-1}$. The black solid contours represent positive values and the red dashed contours represent negative values. The grey ellipse in the bottom left corner represents the full-width at half-maximum (FWHM) of the restoring beam, which is $13\times4.52$\,mas at a position angle $-8.58^\circ$.} \label{fig:vlba}
\end{figure}

The target source's data were exported into DIFMAP \citep{1997ASPC..125...77S} for imaging and model-fitting. The final image was created with natural weighting through the task `CLEAN', see Figure \ref{fig:vlba}. We estimate flux density uncertainties following the instructions described by \citet{1999ASPC..180..301F}. In this work, the integrated flux densities $S_i$ were extracted from Gaussian model-fit in DIFMAP with the task `MODELFIT', where a standard deviation in model-fit was estimated for each component and considered as the fitting noise error. Additionally, we assign the standard 5\% errors originating from amplitude calibration of VLBA (see VLBA Observational Status Summary 2021A \footnote{\url{https://science.nrao.edu/facilities/vlba/docs/manuals/oss2021A}}). The final image gives a $1-\sigma$ rms noise of 0.03\,mJy\,beam$^{-1}$ and a beam size of $13\times4.52$\,mas at a position angle of $-8.58^\circ$.

Furthermore, we also retrieved MERLIN data from MERLIN archive \footnote{\url{http://www.merlin.ac.uk/archive/acknowledge.html}\\\url{emerlin.support@jb.man.ac.uk}} and VLA data from the NRAO data archive \footnote{\url{https://archive.nrao.edu/archive/advquery.jsp}}. The MERLIN 5\,GHz data was observed on June 12th, 2004 and it is likely the same observation that was published by \citet{2006A&A...451...71F}, the total bandwidth of MERLIN observation is 15\,MHz. In this work, we used the reduced data from the MERLIN archive. A manual deconvolution and two-dimensional Gaussian model-fitting were done in DIFMAP with the DIFMAP task `CLEAN' and `MODELFIT', respectively. The final image gives a $1-\sigma$ rms noise of 0.1\,mJy\,beam$^{-1}$ and a beam size of $79\times49$\,mas at a position angle of $26^\circ$. The radio observation at 4.86 GHz and 8.46 GHz were taken from VLA archive project AS0806 (PIs or Observers are Schmitt et al.), which were carried out with a configuration of VLA A-array and a total bandwidth of 50\,MHz, on 2004 October 23. We obtained the reduced data from the NRAO archive and performed a manual deconvolution and two-dimensional Gaussian model-fitting in DIFMAP to retrieve the information of a central compact component, the results are shown in Table \ref{tab:archive}, where the $1-\sigma$ rms noise is 0.06 and 0.04\,mJy\,beam$^{-1}$, respectively. The flux density uncertainties for MERLIN and VLA data are estimated following the same method described in \citet{2020ApJ...904..200Y}.

\section{Results} \label{sec:results}
\begin{figure*}
\centering \includegraphics[scale=0.5]{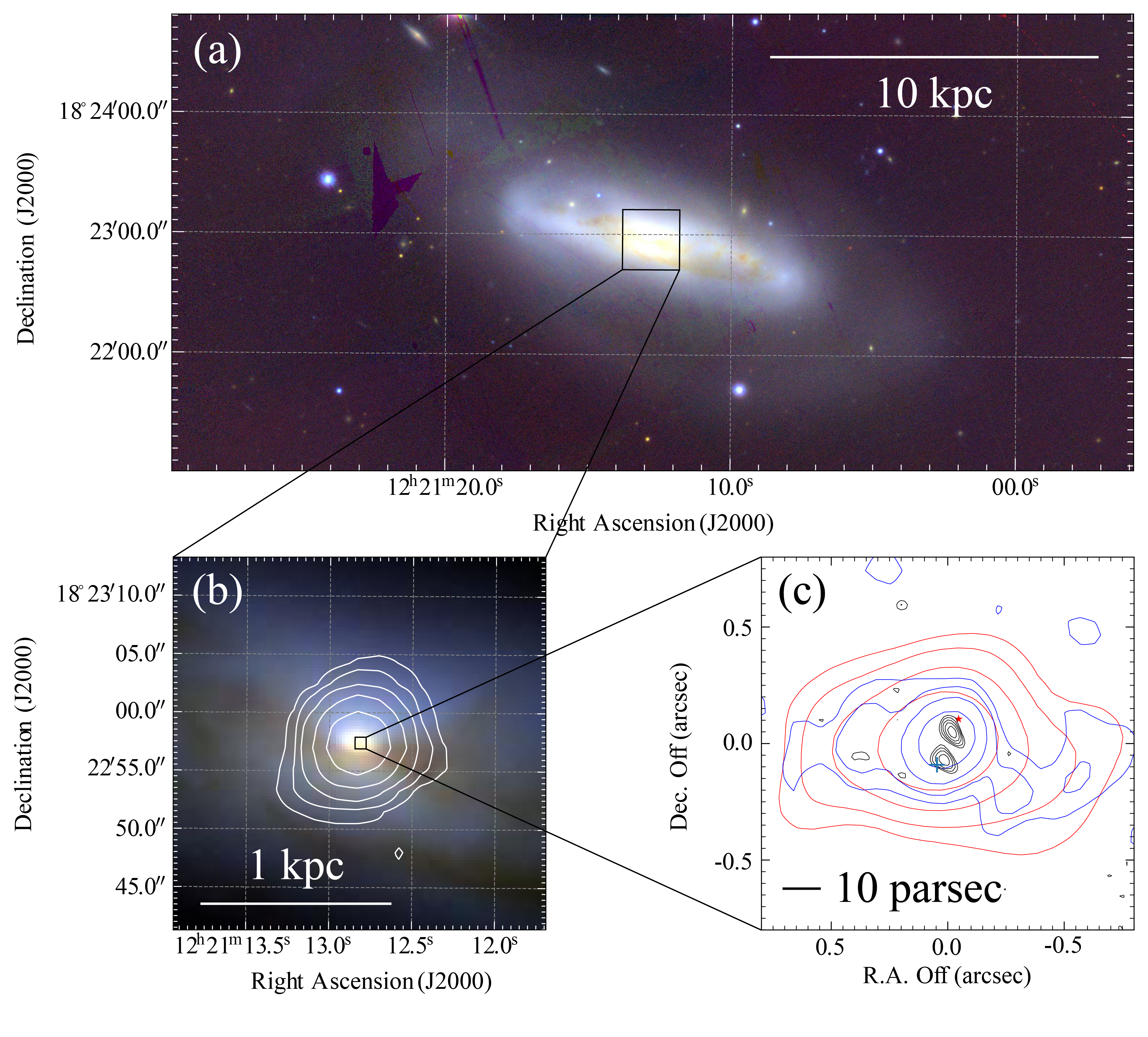}
\caption{A zooming-in to the central engine of \obj. (a) optical image from Pan-STARRS DR1 (PS1); (b) the FIRST 1.4\,GHz contours overlaid on the PS1 color map; (c) radio contours from VLA 4.86\,GHz (red), VLA 8.46\,GHz (blue) and MERLIN 5\,GHz (black) observations, which center on a symmetric center of the VLBA emission in Figure \ref{fig:vlba}, where red star is from \textit{Gaia} early data release 3 (EDR3) and blue cross is from WISE infrared observation. The contours are at 3$\sigma\times(-1, 1, 2, 4, 8,...)$ for VLA 1.4, 4.86 and 8.46\,GHz and 3$\sigma\times(-1, 1, 1.41, 2, 2.83,...)$ for MERLIN 5\,GHz data, here $1\sigma$ noise is $0.13$, $0.06$, $0.04$, and $0.1$\,mJy\,beam$^{-1}$, respectively. \label{fig:full}}
\end{figure*}

In Figure \ref{fig:vlba}, we show the naturally weighted VLBA L-band (1.5 GHz) image. The source shows a symmetric structure with two discrete lobes, where we mark the North-West component as `N' and the South-East component as `S'. The peak flux densities of N and S are $2.70\pm0.17$ and $0.72\pm0.05\,\mathrm{mJy\,beam^{-1}}$, respectively, yielding an SNR of $40$ and $10$, respectively, indicating clear detections. Both components N and S have counterparts in the MERLIN 5\,GHz image, see black contours in panel $c$ of Figure \ref{fig:full}. We performed two-dimensional Gaussian model fittings on the VLBA L-band and MERLIN C-band uv-visibilities. Especially, in order to match the resolution between VLBA L and MERLIN C-band, here we downward the resolution of VLBA L-band data by using an uv-taper. The model-fitting results are listed in Table \ref{tab:results}. We take the integrated flux density to measure the non-simultaneous radio spectral indices $\alpha$ between 1.5 and 5\,GHz for components N and S, which are $-0.64\pm0.09$ and $0.33\pm0.09$, respectively. 

\begin{figure*}
\centering \includegraphics[scale=0.6]{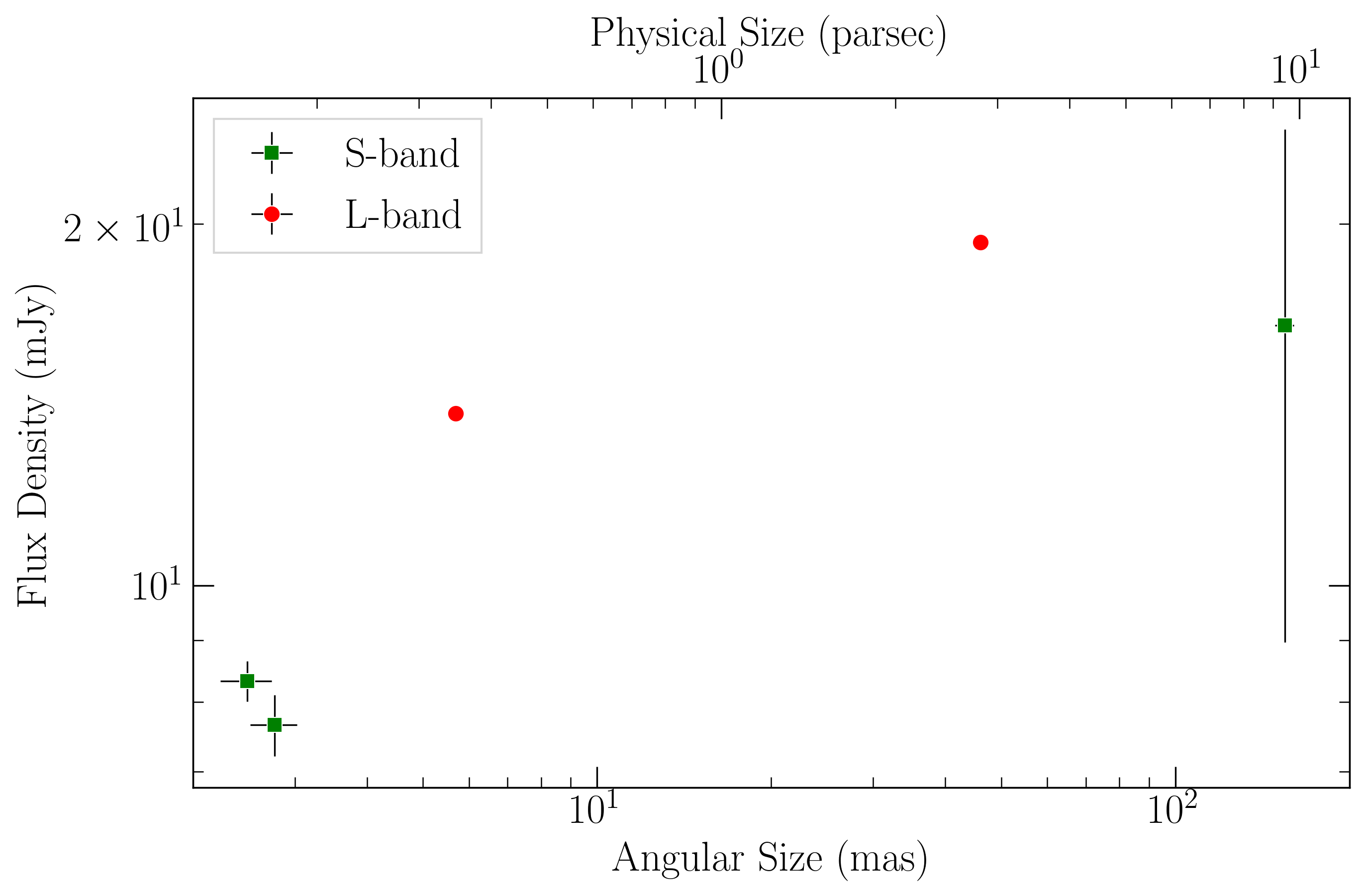}
\caption{The radio flux density of \obj\ over a collection area range from $\sim$2 to $\sim$200\,arcsec. The integrated radio flux densities and uncertainties of \obj\ in L (1.4) and S-band (3\,GHz) are shown (Table \ref{tab:archive}). As \obj\ is not resolved in the given observations, the synthesised beams are taken to represent the collection area. As we use log scales, the intervals between L and S-band flux density represent the radio spectral index.} \label{fig:fs}
\end{figure*}

\begin{figure*}
\centering \includegraphics[scale=0.6]{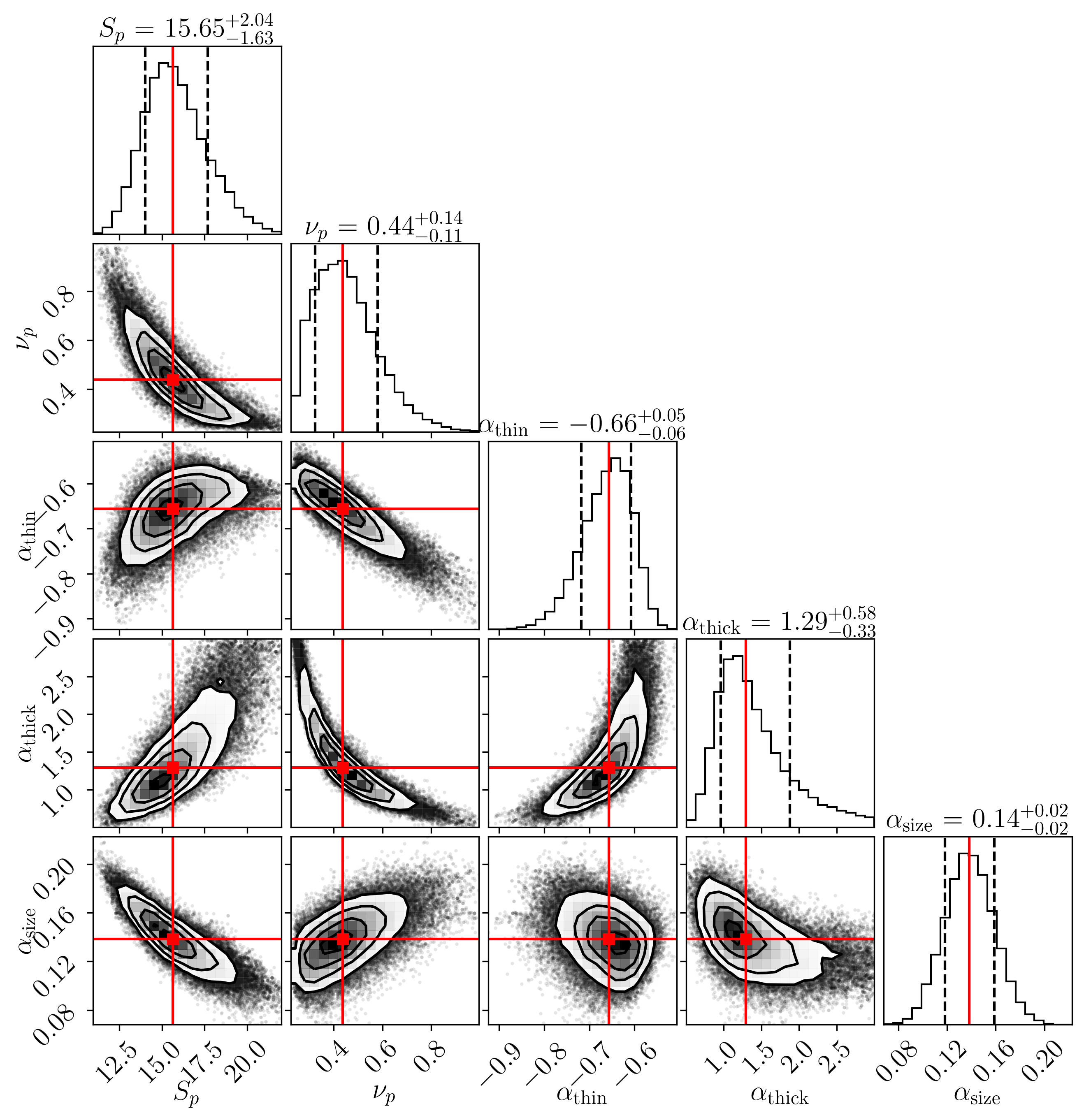}
\caption{Marginalised and joint posterior probability distribution for the model parameters in the MCMC approach. We take 50\% of the distributions as the best-fit values, which are marked in red cross-hairs and red lines. The left and right black dashed lines indicate 16\% and 84\% of the distributions and hence the lower and upper limit for the best-fit values, respectively} \label{fig:mcmc}
\end{figure*}

\begin{figure}
\centering \includegraphics[scale=0.4]{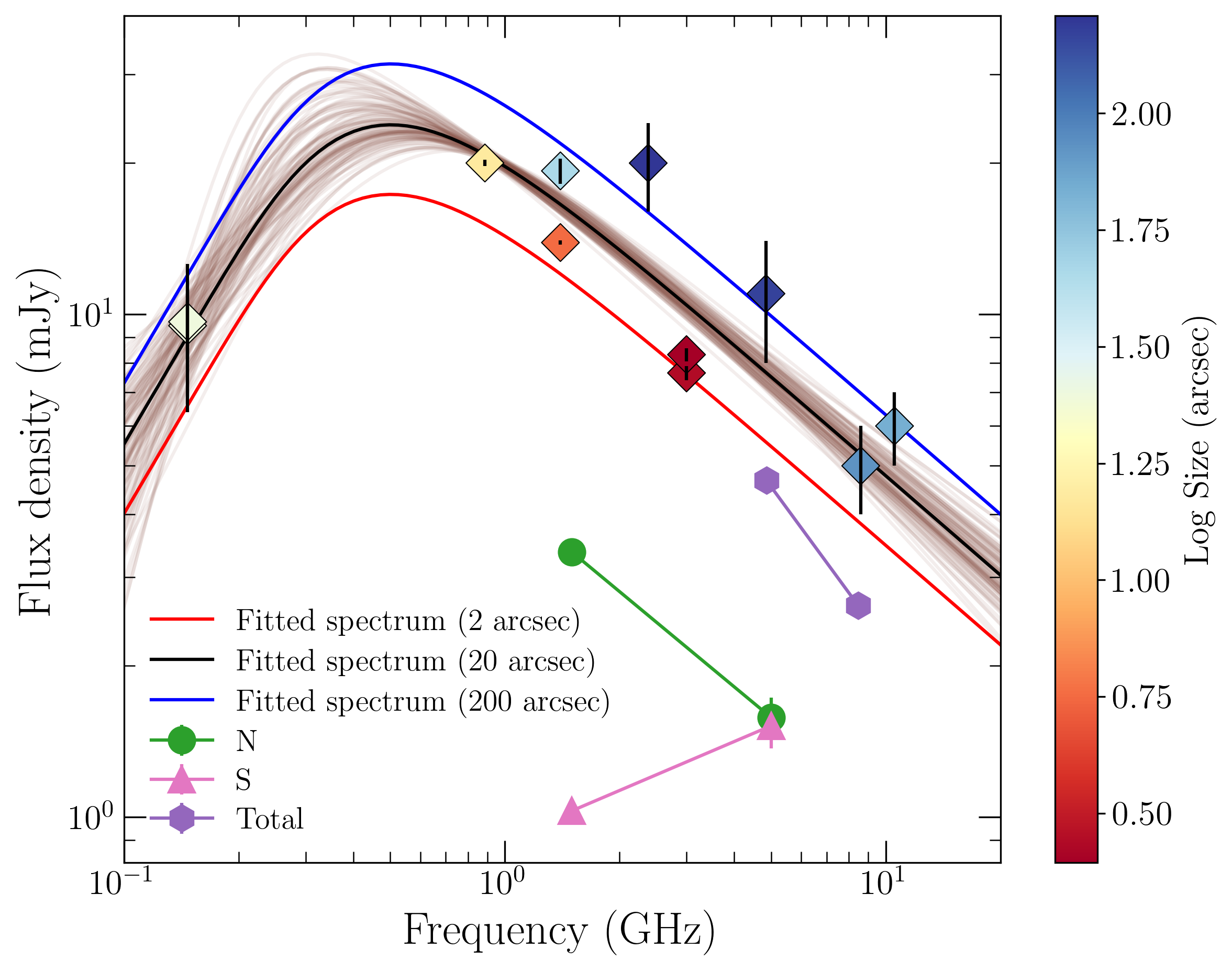}
\caption{Radio spectrum of \obj. The integrated radio flux densities are shown, where the flux densities and uncertainties are taken from Table \ref{tab:archive}. The black, blue, and red solid line is the best-fit spectrum for an angular scale of 20, 200 and 2\,arcsec based on the parameters of the maximum likelihood distribution. The brown lines are from a random sampling in the Markov chain Monte Carlo approach for an angular scale of 20\,arcsec and represent uncertainties of parameters. The green and pink data points are for the components N and S, respectively. The violet data points are for the central compact components in VLA A-array 4.86 and 8.46\,GHz, including both components N and S.}\label{fig:sp}
\end{figure}

By evaluating the radio spectrum compiled from the non-simultaneous data, we check the variability of \obj\ through observations with the same configurations, i.e. at the same frequencies with identical resolutions. Both the two TGSS and VLASS observations satisfy this purpose (see Table \ref{tab:archive}). These two TGSS observations were conducted in the frequency of 0.147\,GHz and an angular scale of 25\,arcsec, which results in no significant variability higher than 35\% within 6 years from 2010 to 2016. Again, these two VLASS observations were conducted in the frequency of 3\,GHz and the angular scale of $\sim$2.6\,arcsec. A marginal variability of 8.7\% can be inferred with a significance of only $2\sigma$ within 2.6\,years from 2019 to 2021. Therefore, as a conclusion, no clear variability is detected in \obj. The target is unresolved at a resolution poorer than 1\,arcsec (see panels $b$ and $c$ of Figure \ref{fig:full}), while combing the high-resolution data with low-resolution ones may still result in the loss of diffuse emission. Figure \ref{fig:fs} shows the L (1.4) and S-band (3\,GHz) observations at different resolutions, where combing the Arecibo 2.38 and Effelsberg 4.85\,GHz data, we measured the 3\,GHz flux density in $\sim$150\,arcsec. The plot implies a positive correlation (possibly a power law) between radio flux densities and collection areas, and it likely hints at a similar trend at different frequencies.

We follow the equation described in e.g. \citet{2017ApJ...836..174C} to model the entire spectrum of \obj. Additionally, here we assume a power-law distribution of radio flux density along angular size according to the observational hints (see Figure \ref{fig:fs}).
\begin{equation}\label{eq:sp}
S_\nu= \frac{S_p\theta^{\alpha_{size}}}{(1-e^{-1})}\times{\left(1-e^{{-(\nu/\nu_p)}^{\alpha_\mathrm{thin}-\alpha_\mathrm{thick}}}\right)}\times{\left(\frac{\nu}{\nu_p}\right)^{\alpha_\mathrm{thick}}}
\end{equation}
where $S_\nu$ is the flux density at frequency $\nu$ and angular size $\theta$, $\nu_p$ is the frequency at which the spectrum peaks, $S_p$ is the flux density at the frequency $\nu_p$ and an unit angular size (i.e. $\theta$=1\,arcsec), $\alpha_\mathrm{thick}$ and $\alpha_\mathrm{thin}$ are the spectral indices in the optically thick and optically thin regimes of the spectrum, respectively.

A Markov chain Monte Carlo algorithm is used to fit the spectrum, and only the observations with beam size larger than 2\,arcsec are taken in the fitting. In Figure \ref{fig:mcmc}, we show the posterior probability distributions of each parameter, and we take 16\% and 84\% of the distributions as the lower and upper limits, respectively, thus representing $1\sigma$ confidence ranges. The fitting yields a peak flux density $S_p=15.65^{+2.04}_{-1.63}$\,mJy at frequency $\nu_p=0.44^{+0.14}_{-0.11}$\,GHz and the angular size $\theta=1$\,arcsec, with optically thin and thick spectral index $\alpha_\mathrm{thin}=-0.66^{+0.05}_{-0.06}$ and $\alpha_\mathrm{thick}=1.29^{+0.58}_{-0.33}$, respectively. Furthermore, the power-law index of flux density distribution along angular size $\alpha_\mathrm{size}=0.14\pm0.02$. The spectrum of \obj\ is plotted in Figure \ref{fig:sp}, where the black solid line represents the best fit of equation \ref{eq:sp} to the integrated radio spectrum at the typical angular size of $20$\,arcsec. The thin and thick radio spectral indices are consistent with the typical peaked spectrum radio sources \citep{1998PASP..110..493O}. Furthermore, \obj\ is resolved and shows diffuse emission along east-west direction in an angular scale of 1\,arcsec (see panel $c$ of Figure \ref{fig:full}). It shows that the spectrum of the central compact component in VLA A-array 4.86 and 8.46\,GHz aligns the integrated spectrum (optically thin part, see Figure \ref{fig:sp}), which indicates the core dominance in this scale. Assuming a similar flux density distribution along frequencies and sizes, i.e. equation \ref{eq:sp} is still applicable in an angular scale of $\sim0.5$\,arcsec, the equation gives a flux density of $4.67\pm1.19$ and $3.10\pm0.85$\,mJy\,beam$^{-1}$ for VLA A-array 4.86 and 8.46\,GHz respectively and this is approximately consistent with the observations.

The brightness and compactness of radio components can be parameterized by taking the brightness temperatures, measured using the formula \citep[e.g.][]{2005ApJ...621..123U} 
\begin{equation}\label{eq:bt}
T_\mathrm{B}=1.8\times10^9(1+z)\frac{S_i}{\nu^2\theta^2}~\mathrm{(K)},
\end{equation}
where \(S_i\) is the integrated flux density of each Gaussian model component in mJy (column 2 of Table \ref{tab:results}), $\theta$ is $\mathrm{FWHM}$ of the Gaussian model in mas (column 4 of Table \ref{tab:results}), \(\nu\) is the observing frequency in GHz, and \(z\) is the redshift. The estimated brightness temperatures are listed in column 5 of Table \ref{tab:results}. The radio brightness temperatures are inversely proportional to the measured component sizes. Since the measured component sizes are only upper limits, the radio brightness temperatures should be considered as lower limits.

Assuming minimum energy (approximately equipartition) conditions in synchrotron emission, we can estimate the magnetic field $B_\mathrm{min}$ in Gauss (G) through the formula \citep[e.g.][]{2022ApJ...934...26P}
\begin{equation}
B_\mathrm{min}\approx0.0152\left[\frac{a}{f_{rl}}\frac{(1+z)^{4-\alpha}}{\theta^3}\frac{S_i}{\nu^\alpha}\frac{\nu_2^{p+\alpha}-\nu_1^{p+\alpha}}{r(p+\alpha)}\right]^{2/7}~\mathrm{(G)}
\end{equation}
$S_i$ (in mJy) is the integrated flux density of the source measured at frequency $\nu$ (in GHz) and angular size $\theta$ (in mas), $\alpha$ is the spectral index, $z$ is the redshift of the source, and $r$ is the comoving distance in Mpc. Here we take $p=0.5$, $\alpha=\alpha_\mathrm{thin}$ in the optically thin part of the spectrum from $\nu_1 \sim1$\,GHz to $\nu_2\sim10$\,GHz, a filling factor for the relativistic plasma $f_{rl} = 1$, a relative contribution of the ions to the energy a = 2, and the comoving distance $r$ in Mpc. By adopting the standard $\Lambda$CDM cosmology and using the cosmology calculator provided by NED \footnote{\url{https://www.astro.ucla.edu/~wright/CosmoCalc.html}}, $r=12.7\,\mathrm{Mpc}$ in this case. Here we use the integrated flux density of 19.83\,mJy estimated through formula \ref{eq:sp} and fitting parameters, at the typical frequency of 1\,GHz and angular size of 20\,arcsec, which yield $B_\mathrm{min}=3\times10^{-4}\,\mathrm{G}$. The magnetic field is consistent with the typical value of PS sources, e.g. gigahertz peaked-spectrum (GPS, $B\sim10^{-3}\,\mathrm{G}$) and compact steep spectrum (CSS, $B\sim10^{-4}\,\mathrm{G}$) radio sources \citep{1998PASP..110..493O}.

The electron lifetime for a turnover-type spectral radio source can be estimated as follow
\citep{1998PASP..110..493O}
\begin{equation}
t\simeq8.22\times\frac{B^{1/2}}{B^2+B_{R}^2}\times{\left((1+z)\nu_p\right)}^{-1/2}~\mathrm{(yr)}
\end{equation}
where $B$ is the magnetic field in $\mathrm{G}$, $B_\mathrm{R}\simeq4(1+z)^2\times10^{-6}\,\mathrm{G}$ is the equivalent magnetic field of the microwave background, and $\nu_p$ is the break frequency in GHz. For \obj, using these values estimated above, i.e. $B=3\times10^{-4}\,\mathrm{G}$ and $\nu_p=0.44\,\mathrm{GHz}$, we find the electron lifetime is $\sim1\times10^6$ years.

\begin{table*}
\centering
\caption{Total radio flux density from archival observations.}
\label{tab:archive}
\begin{threeparttable}
\begin{tabular}{cccccc}
\hline
\hline
Survey or Facility & Frequency & Date & $S_i$   & Beam & References \\
                   & (GHz)     &      & (mJy)   & (arcsec)  & \\
\hline
TGSS                &   0.147    &2016-03-15&$    9.50\pm3.11 $&$  25  $& \citet{2017AA...598A..78I}  \\
TGSS                &   0.147    &2010-05-27&$    9.67\pm1.53 $&$  25  $& \citet{2017AA...598A..78I}  \\
RACS                &   0.887    &2019-04-24&$   20.01\pm0.29 $&$  15  $& \citet{2020PASA...37...48M}  \\
NVSS                &   1.4      &1993-11-15&$   19.3\pm1.1   $&$  46  $& \citet{1998AJ....115.1693C} \\
FIRST         &   1.4      &1999-01-15&$   13.90\pm0.13 $&$  5.69\pm0.17  $& \citet{1997ApJ...475..479W} \\
Arecibo             &   2.38     &1975-08-01&$   20\pm4       $&$ 162  $& \citet{1978ApJS...36...53D} \\
VLASS         &   3        &2019-03-21&$   7.65\pm0.25  $&$  2.77\pm0.44  $& \citet{2020PASP..132c5001L} \\
VLASS         &   3        &2021-11-20&$   8.32\pm0.25  $&$  2.48\pm0.32  $& \citet{2020PASP..132c5001L} \\
Effelsberg          &   4.85     &2000-2003&$    11\pm3       $&$ 147  $& \citet{2004AA...418....1V} \\
Effelsberg          &   8.6      &2000-2003&$     5\pm1       $&$  85  $& \citet{2004AA...418....1V} \\
Effelsberg          &   10.55    &2000-2003&$    6\pm1        $&$  69  $& \citet{2004AA...418....1V} \\
VLA-A$^\dagger$     &   4.86       &2004-10-23&$    4.68\pm0.10 $&$  0.58\pm0.08 $& This work \\
VLA-A$^\dagger$     &   8.46       &2004-10-23&$    2.64\pm0.06 $&$  0.41\pm0.05 $& This work \\
\hline
\end{tabular}
\begin{tablenotes}
\small
\item $\dagger$: Model-fitting results of the bright and compact nuclear component. \\
\item TGSS: the TIFR GMRT Sky Survey; RACS: the Rapid ASKAP Continuum Survey; NVSS: the NRAO VLA Sky Survey; FIRST: Faint Images of the Radio Sky at Twenty-centimeters; VLASS: the Very Large Array Sky Survey. 
\end{tablenotes}
\end{threeparttable}
\end{table*}

\begin{table*}
\caption{Model-fitting results of components N and S in \obj}
\label{tab:results}
\begin{threeparttable}
\begin{tabular}{ccccccccc}
\hline
\hline
Components &  $S_i$  &  $S_p$             & $\phi$ & $\log{T_\mathrm{B}}$ & $\log{L_\mathrm{R}}$ &$\theta_{b,maj}$ &$\theta_{b,min}$  & PA         \\
&  (mJy)  &  (mJy\,beam$^{-1}$)&(mas)&(K)&(erg/s)&(mas)&(mas)&($\circ$) \\
\hline
\multicolumn{9}{c}{VLBA 1.5\,GHz (Tapered)} \\
\hline
N                &$3.37\pm0.18$&$2.70\pm0.17$&$ 12.5$&$  7.2 $&$ 36.00\pm0.02 $&$ 33 $&$ 18 $&$ 44 $\\
S                &$1.03\pm0.06$&$0.72\pm0.05$&$ 15.0$&$  6.5 $&$ 35.48\pm0.02 $&$ 33 $&$ 18 $&$ 44 $\\
\hline
\multicolumn{9}{c}{MERLIN 5\,GHz} \\
\hline
N                &$1.58\pm0.15$&$1.51\pm0.18$&$ 6.5$&$  6.4 $&$ 36.18\pm0.04 $&$ 79  $&$ 49 $&$ 26  $\\
S                &$1.52\pm0.15$&$1.08\pm0.18$&$ 37.9$&$  4.8 $&$ 36.16\pm0.04 $&$ 79  $&$ 49 $&$ 26  $\\
\hline
\end{tabular}

\begin{tablenotes}
\item Column 1: component name; Column 2: integrated flux density; Column 3: peak flux density; Column 4: deconvolved component size; Column 5: radio brightness temperature; Column 6: total luminosity; Column 7 - 9: beam major axis, minor axis, and position angle.
\end{tablenotes}

\end{threeparttable}
\end{table*}

\section{Discussion} \label{sec:dis}

\subsection{A peaked spectrum radio source} \label{subsec:cso}
The brightness temperatures for components N and S are $10^{7.2}$ and $10^{6.5}$\,K, respectively, favoring a non-thermal origin. Both components N and S show edge-brightening structure and inverse symmetric appearance with each other (see Figure \ref{fig:vlba}). The angular distance between N and S is $\sim120$\,mas, corresponding to $\sim7$\,parsec. The whole milliarcsec-scale radio morphology of \obj\ resembles a compact symmetric object \citep[CSO,][]{1998PASP..110..493O,2021A&ARv..29....3O}. Along with the radio spectrum and 1.4\,GHz radio power of $\sim10^{20}\,\mathrm{W\,Hz^{-1}}$, the source belongs to a sample of (megahertz) peaked spectrum (PS/MPS) radio sources \citep{2021A&ARv..29....3O} \citep[or based on the historical definition, \obj\ can be marginally identified as gigahertz peaked-spectrum or compact steep-spectrum radio sources,][]{1998PASP..110..493O}. The PS scenario hints at an absorbed radio core in the symmetric center of the whole structure. The integrated radio spectrum of the source shows a turnover at the frequency 0.44\,GHz, which is consistent with a low redshift PS source. On the other hand, the whole morphology and integrated spectrum are unlikely to be caused by a core-jet structure. Again, the core-jet scenario, where component S (inverted spectrum, see below) is the core and component N (steep spectrum, see below) is a radio lobe, will result in a discrepancy: the non-detection of a counter-jet lobe indicates that jet versus counter-jet flux ratio $>30$ in this scenario, while explaining this as a beaming effect requires large advancing speed and small view angle to the jet beam. It is unlikely that the object acts this way due to the extremely low advancing speed (see below), the edge-on disk of the host and the nature that the source is not a blazar. Therefore, it again disfavors the core-jet explanation. On the origin of the spectral turnover, the rising spectral index $\alpha_\mathrm{thick}=1.29^{+0.58}_{-0.33}$ is below the critical slope $\alpha_c=+2.5$ of a free-free absorption. Thus, it favors an origin of synchrotron self-absorption mechanism, while a free-free absorption cannot be fully ruled out with the limited data \citep[][]{2021A&ARv..29....3O}.  Blackhole mass measurements for young radio sources (including PS sources) indicate that they have a mass range from $\log{(M_\mathrm{BH}/M_\odot)}\sim7.32$ to $9.84$ \citep{2020MNRAS.491...92L}. Therefore, whatever the different black hole mass measurements of \obj, the source stands out as one of the lightest and nearest PS sources until now \citep[e.g.][]{2004MNRAS.348..227S,2021A&ARv..29....3O}.

Interestingly, the brightness of components N and S is quite asymmetric (see Figure \ref{fig:vlba}), and their VLBA 1.5\,GHz luminosity ratio is $3.23$, hinting at an interaction with dense medium \citep[see, e.g.][]{2001MNRAS.321...37S, 2003MNRAS.341...91T, 2021A&ARv..29....3O}. Additionally, the lobes of components N and S have lateral structures other than pointing directly away from the symmetric center (or the nucleus), further suggesting the deflection of a jet due to a dense inhomogeneous and asymmetric external medium on opposite sides of the nucleus \citep[e.g.][]{2007ApJ...655..769C, 2021A&ARv..29....3O}. According to the spectral age (electron lifetime) of $\sim1\times10^6$ years, the source was likely frustrated by a dense medium as it results in an extremely low advancing velocity ($1.1\times10^{-5}\,c$ or $3.4\,\mathrm{km\,s^{-1}}$, where $c$ is the speed of light) for the lobes. Indeed, dense molecular gas was observed in \obj\ and concentrated in the nuclear region \citep{2002PASJ...54..555K}.

The $1.5$ - $5$\,GHz radio spectrum of component N is aligned with the integrated spectrum, while component S has a deviation from the integrated spectrum. A natural assumption is that both components are produced simultaneously, therefore having similar internal parameters. If so, the inverted spectrum at the location of component S requires an additional mechanism to further absorb the low-energy electrons -- the free-free absorption by the external ionized dense gas is possibly at work for the southern component. The exactly different spectral indices for the jet and the counter-jet lobes are common in radio sources \citep[e.g.][]{2000ApJ...530..233W, 2011A&A...535A..24S} and one of the promising explanations is the free-free absorption by circumnuclear medium \citep[e.g.][]{2000ApJ...530..233W}. Again the hypothesis is supported by the Wide-field Infrared Survey Explorer (WISE) emission near the position of component S. The ALLWISE program obtained $1\sigma$ astrometric accuracy of $0.033$\,arcsec \citep{2014yCat.2328....0C} (see error-bars in panel c of Figure \ref{fig:full}), therefore, it is still close to component S by taking $3\sigma$ positional uncertainty of $\sim0.1$\,arcsec and clearly differs from component N. $Gaia$ Early Data Release 3 (EDR3) measured the optical core of \obj\ and obtained $3\sigma$ astrometric uncertainty of $\sim0.01$\,arcsec \citep{2016A&A...595A...1G, 2021A&A...649A...1G}, which is close to component N. Because the host galaxy is edge-on, the optical position naturally offsets towards the north direction (see panels $a$ and $b$ of Figure \ref{fig:full}) due to the obstruction of the dust lane.

\subsection{On the black hole mass and accretion}

In measuring black hole mass, both the stellar velocity and the $K_s$-band bulge luminosity-based methods have intrinsic dispersion of at least 1 dex \citep{2013ARA&A..51..511K}, which is even worse in the low-mass end. On the other hand, LINERs have very different accretion flow \citep{2003ApJ...583..159H, 2008ARA&A..46..475H, 2009ApJ...699..626H} and broad emission line regions \citep[e.g.][]{1997ApJS..112..391H}. Therefore, estimating black hole mass from the single-epoch broad H$\alpha$ emission for the LINER \obj\ \citep{2018ApJ...863....1C} may induce a large uncertainty. Again, LINERs tend to have extremely low accretion rates \citep{2003ApJ...583..159H, 2008ARA&A..46..475H, 2009ApJ...699..626H}, while the unreasonably high Eddington ratio $\lambda_\mathrm{Edd}=0.25$ ($>0.01$) \citep{2018ApJ...863....1C} would further hint at an underestimation of black hole mass in \citet{2018ApJ...863....1C}. It is essential to constrain the black hole mass of \obj\ with a radio-based method, e.g. the fundamental plane of black hole activity \citep{2003MNRAS.345.1057M}.

The fundamental plane relation among nuclear radio luminosity, nuclear X-ray luminosity, and black hole mass unified the accretion and ejection process in compact systems. The existence of such a relationship is based on that the radio emission is produced in a jet/outflow, and X-ray emission is produced in a disk-corona system. Both radio and X-ray power are related to black hole mass and accretion rate. Therefore, the fundamental plane relation is thought to work in any accretion system which is in a quiescent or low/hard accretion state \citep[associated with a steady ejection, see][]{2003MNRAS.344...60G, 2003MNRAS.345.1057M}. Additionally, the very high/intermediate state may also produce radio ejection that can follow the same trend \citep[see][they also include transient sources]{2003MNRAS.345.1057M}. However, including sources in a very high/intermediate state induces a dispersion in the fundamental plane relation. This is primarily due to the evolution of individual radio blobs, as the radio ejecting process is episodic in this state. There are several works exploring the fundamental plane relation on CSOs \citep[morphological candidates of PS sources,][]{2016ApJ...818..185F, 2020ApJ...892..116W, 2020MNRAS.497..482L}. Most of the results suggest CSOs do deviate from the classical trends. Two types of known contamination are (1) radio emissions from lobes will be enhanced when they propagate through a dense medium \citep{2021A&ARv..29....3O}; (2) X-ray emission contains a contribution from the jet, e.g., through synchrotron or inverse Compton mechanisms \citep{2008ApJ...680..911S}. Furthermore, the radio emissions from lobes of CSOs are not associated with the X-ray emission collected in the core region, because the radio emissions from lobes are substantially produced in different epochs from the core X-ray emission, e.g. $\sim1\times10^6$\,years ago in \obj.

\obj\ have Eddington ratio $\lambda_\mathrm{Edd}=0.001-0.25$, corresponding to the black hole masses range from $10^{7.7}$ to $10^{5.3}\,M_\odot$. The Eddington ratio $\lambda_\mathrm{Edd}{\equiv}L_\mathrm{bol}/L_\mathrm{Edd}$, where $L_\mathrm{bol}=10L_\mathrm{B}$ and $L_\mathrm{Edd}=1.26\times10^{38}(M_\mathrm{BH}/M_\odot)$\,(erg/s) \citep[see also][]{2020ApJ...904..200Y}. The calculation above takes $10L_\mathrm{B}$ as the bolometric luminosity, where $L_\mathrm{B}=41.8\,\mathrm{erg\,s^{-1}}$ is $B$-band luminosity of \obj. Here the $B$-band luminosity is estimated from H$\alpha$ line luminosity of \obj\ \citep{2018ApJ...863....1C} through the method described in \citep[e.g.][]{2007ApJ...670...92G, 2012ApJ...755..167D, 2018ApJ...863....1C}. The critical Eddington ratio from low to high-state is $\lambda_\mathrm{Edd}=0.01$ \citep{2003MNRAS.344...60G}. Furthermore, a recent work (Yang et al. In preparation) studies a sample of IMBHs with VLBA, and it indicates that parsec-scale radio emissions from IMBHs can follow \citet{2003MNRAS.345.1057M}'s fundamental plane very well. In order to keep the simultaneity between radio and X-ray observations, here we firstly estimate the black hole mass of \obj\ by taking the radio emission from the core region. If we take a $3\sigma=0.3$\,mJy upper limit in the MERLIN image as the core radio flux density, then the 5\,GHz core radio luminosity is $\log{L_\mathrm{core, 5GHz}}=35.46$\,erg\,s$^{-1}$ (redshift effect can be negligible). Here we take the X-ray luminosity from the $Chandra$ observation and assume that $\log{L_\mathrm{X(2-10keV)}}\simeq\log{L_\mathrm{X(0.5-8keV)}}=39.69\,\mathrm{erg\,s^{-1}}$ \citep{2022MNRAS.512.3284S}. Additionally, an observation with $ROSAT$ gives $\log{L_\mathrm{X(0.1-2.4keV)}}=39.48\,\mathrm{erg\,s^{-1}}$ \citep{2001AJ....122..637H}. The resulting black hole mass of \obj\ is $\log{(M_\mathrm{BH}/M_\odot)}=5.53$, consistent with the latest measurement by \citet{2018ApJ...863....1C}. An exploration of fundamental plane relation on a sample of CSOs (with radio flux density from lobes) indicates that they can follow the trend, while their radio luminosity is $\sim$1 dex higher than the original fitting of the fundamental plane relation \citep[see][]{2020ApJ...892..116W}. If we consider the 1 dex deviation and use the radio luminosities of lobes, a similar black hole mass $\log{(M_\mathrm{BH}/M_\odot)}\sim5.16$ can be estimated. As most of CSOs have a radio-quiet core \citep{2021A&ARv..29....3O}, the use of the lobe's radio luminosity in the fundamental plane should be explored further. We also used the method described here to measure the black hole mass of another IMBH candidate RGG~9 \citep{2013ApJ...775..116R}. Using the radio luminosity presented by \citet{2020MNRAS.495L..71Y}, our black hole mass estimate matches very well with the measurement by using virial techniques \citep{2013ApJ...775..116R}.

We have to bear in mind that \citet{2003MNRAS.345.1057M}'s fundamental plane of black hole activity has a large dispersion. Taking the upper limit of the 5\,GHz core radio luminosity, we estimate black hole mass using \citet{2018A&A...616A.152S} correlation as $\log{(M_\mathrm{BH}/M_\odot)}=5.9\pm2.0$. Again, using the most recent version of the fundamental plane of BH activity \citep{2019ApJ...871...80G}, we obtain $\log{(M_\mathrm{BH}/M_\odot)}=5.9\pm0.3$. Alternatively, directly using the 1 dex reduced radio luminosity of lobes and \citet{2019ApJ...871...80G} correlation, we roughly obtain $\log{(M_\mathrm{BH}/M_\odot)}=5.6\pm0.3$. Furthermore, we note that the X-ray luminosity of \obj\ is under-estimated due to the absorption of the circumnuclear medium in the host galaxy \citep{2001AJ....122..637H, 2022MNRAS.512.3284S}, therefore resulting in an over-estimating of black hole mass through the fundamental plane of black hole activity. In summary, the fundamental plane of black hole activity would refine the black hole mass of \obj\ to be $\lesssim10^6\,M_\odot$. Therefore, the object is a low-mass AGN and a potential candidate for IMBHs as well.

In X-ray binaries, an episodic ejection is produced when the accretion is in a very high/intermediate state \citep{2004MNRAS.355.1105F}. Interestingly, an episodic ejection is already observed in the IMBH candidate HLX-1 when it is in a high X-ray state \citep{2012Sci...337..554W}. In the unification model of accretion process \citep{2006Natur.444..730M}, AGNs should also have specific accretion states and relevant state transitions as that are universally observed in Galactic XRBs \citep{2004MNRAS.355.1105F}. Naturally, multi-types of accretion state should also be expected in IMBHs, and according to the scaling relation, it should play much faster than it is in AGNs \citep[as the timescale is proportional to black hole masses, e.g.][]{2017A&A...603A.127S}. Here, the spectral age of the radio emission in \obj, $\sim1\times10^6$\,years, is slightly larger than an episodic activity of the radiation pressure instabilities in accretion disks \citep[i.e. $10^4-10^5$\,year,][]{2021A&ARv..29....3O}. If so, future high-resolution observations of IMBHs at low frequencies ($<1\,$GHz) will help to reveal radio emissions or bubbles from the relic of ejecta. Monitoring observations may also help in capturing the episodic ejection directly and revealing the evolution of radio blobs.

\section{Conclusion} \label{sec:conclusions}
In summary, we performed a Very Long Baseline Array (VLBA) L-band (1.5 GHz) observation of the IMBH candidate in the nucleus of the LINER galaxy NGC~4293 ($z=0.003$). By cooperating with the archive VLA and MERLIN data, we find (1) the source has two radio blobs with a projected distance of $\sim7$\,parsec apart from each other and its integrated spectrum has a turnover at $\sim0.44$\,GHz. Therefore it can be allocated as a typical compact symmetric object (CSO). We also find evidence that the lobes are interacting with the dense medium. Despite different mass estimations, the source stands out as one of the lightest and nearest CSOs; (2) We estimate the black hole mass of \obj\ through the fundamental plane of black hole activity and constrain it to be $\lesssim10^6\,M_\odot$. The result supports that \obj\ has a less massive AGN; (3) The discovery of compact symmetric ejecta in the less massive (or intermediate-mass) AGN, \obj, can be reasonably explained as the source is maintaining an episodic ejection.

\section*{Acknowledgements}

X.-L.Y. designed the VLBA observations, made the VLBI data reduction and model-fitting, and interpret the results, R.-L.W. contribute to drafting the manuscript, interpret the results and proof correction, and Q.G. provided useful comments on the manuscript. This work is supported by the Shanghai Sailing Program (21YF1455300) and China Postdoctoral Science Foundation (2021M693267). XLY thanks for the support from the National Science Foundation of China (12103076). We thank Luis C. Ho from the Kavli Institute for Astronomy and Astrophysics at Peking University (KIAA-PKU), he shared important experiences and matters that need attention in measuring black hole masses of LINER galaxies.
The National Radio Astronomy Observatory is a facility of the National Science Foundation operated under cooperative agreement by Associated Universities, Inc. 
$e$-MERLIN is a National Facility operated by the University of Manchester at Jodrell Bank Observatory on behalf of STFC.
This work has made use of data from the Pan-STARRS1 Surveys (PS1) and the PS1 public science archive. 
This work has made use of data from the European Space Agency (ESA) mission {\it Gaia} (\url{https://www.cosmos.esa.int/gaia}), processed by the {\it Gaia} Data Processing and Analysis Consortium (DPAC, \url{https://www.cosmos.esa.int/web/gaia/dpac/consortium}). 

\section*{Data Availability}
Scientific results from data presented in this publication are derived from the VLBA project BA146. The correlated data are available in the NRAO Science Data Archive (\url{https://data.nrao.edu}).



\bibliographystyle{mnras}
\bibliography{imbh_cso} 







\bsp	
\label{lastpage}
\end{document}